\documentclass[twocolumn,twocolappendix]{aastex63}

\usepackage{soul}
\usepackage{color}
\usepackage{hyperref}

\shorttitle{Indication of a pulsar wind nebula in SN 1987A}
\shortauthors{Greco et al.}

\graphicspath{{./}{figures/}}

\begin{document}

\title{Indication of a Pulsar Wind Nebula in the hard X-ray emission from SN 1987A}

\correspondingauthor{Emanuele Greco}
\email{emanuele.greco@inaf.it}

\author[0000-0001-5792-0690]{Emanuele Greco}
\affiliation{Universit\`a degli Studi di Palermo, Dipartimento di Fisica e Chimica, Piazza del Parlamento 1, 90134 Palermo, Italy}
\affiliation{INAF-Osservatorio Astronomico di Palermo, Piazza del Parlamento 1, 90134 Palermo, Italy}

\author{Marco Miceli}
\affiliation{Universit\`a degli Studi di Palermo, Dipartimento di Fisica e Chimica, Piazza del Parlamento 1, 90134 Palermo, Italy}
\affiliation{INAF-Osservatorio Astronomico di Palermo, Piazza del Parlamento 1, 90134 Palermo, Italy}

\author{Salvatore Orlando}
\affiliation{INAF-Osservatorio Astronomico di Palermo, Piazza del Parlamento 1, 90134 Palermo, Italy}

\author{Barbara Olmi}
\affiliation{INAF-Osservatorio Astronomico di Palermo, Piazza del Parlamento 1, 90134 Palermo, Italy}

\author{Fabrizio Bocchino}
\affiliation{INAF-Osservatorio Astronomico di Palermo, Piazza del Parlamento 1, 90134 Palermo, Italy}

\author{Shigehiro Nagataki}
\affiliation{Astrophysical Big Bang Laboratory (ABBL), RIKEN Cluster for Pioneering Research, 2-1 Hirosawa, Wako, Saitama 351-0198, Japan}
\affiliation{RIKEN Interdisciplinary Theoretical and Mathematical Science Program (iTHEMS), 2-1 Hirosawa, Wako, Saitama 351-0198, Japan}

\author{Masaomi Ono}
\affiliation{Astrophysical Big Bang Laboratory (ABBL), RIKEN Cluster for Pioneering Research, 2-1 Hirosawa, Wako, Saitama 351-0198, Japan}
\affiliation{RIKEN Interdisciplinary Theoretical and Mathematical Science Program (iTHEMS), 2-1 Hirosawa, Wako, Saitama 351-0198, Japan}

\author{Akira Dohi}
\affiliation{Department of Physics, Kyushu University, 744 Motooka, Nishi-Ku, Fukuoka Fukuoka 819-0395, Japan}

\author{Giovanni peres}
\affiliation{Universit\`a degli Studi di Palermo, Dipartimento di Fisica e Chimica, Piazza del Parlamento 1, 90134 Palermo, Italy}
\affiliation{INAF-Osservatorio Astronomico di Palermo, Piazza del Parlamento 1, 90134 Palermo, Italy}
\begin{abstract}

Since the day of its explosion, SN 1987A (SN87A) was closely monitored with the aim to study its evolution and to detect its central compact relic. The detection of neutrinos from the supernova strongly supports the formation of a neutron star (NS). However, the constant and fruitless search for this object has led to different hypotheses on its nature. Up to date, the detection in the ALMA data of a feature somehow compatible with the emission arising from a proto Pulsar Wind Nebula (PWN) is the only hint of the existence of such elusive compact object. Here we tackle this 33-years old issue by analyzing archived observations of SN87A performed by {\it Chandra} and {\it NuSTAR} in different years. We firmly detect nonthermal emission in the $10-20$ kev energy band, due to synchrotron radiation. The possible physical mechanism powering such emission is twofold: diffusive shock acceleration (DSA) or emission arising from an absorbed PWN. By relating a state-of-the-art magneto-hydrodynamic simulation of SN87A to the actual data, we reconstruct the absorption pattern of the PWN embedded in the remnant and surrounded by cold ejecta. We found that, even though the DSA scenario cannot be firmly excluded, the most likely scenario that well explains the data is the PWN emission.  
\end{abstract}


\section{Introduction} 
\label{sec:intro}

SN 1987A (SN87A) in the Large Magellanic Cloud (LMC) was a hydrogen-rich core-collapse supernova (SN) discovered on 1987 February 23 \citep{wls87}. It occurred approximately 51.4 kpc from Earth \citep{pan99} and its dynamical evolution is strictly related to the very inhomogenous circumstellar medium (CSM), composed by a dense ring-like structure within a diffuse HII region \citep{sck05}. SN87A is the first naked-eye SN exploded since telescopes exist and its evolution has been deeply monitored in various wavelengths \citep{mcr93,mcf16}. In particular, the X-ray band is ideal to investigate the interaction of the shock front with the CSM and the emission of the expected central compact leftover of the supernova explosion.

Despite the unique consideration granted with deep and continuous observations, and the neutrinos detection \citep{bbb87} strongly indicating the formation of a neutron star \citep{vis15}, the elusive compact object of SN87A is still undetected. The most likely explanation for this non-detection is ascribable to the absorption due to ejecta, i.e. the dense and cold material ejected by the supernova \citep{fcr87}: because of the young age of SN87A, the ejecta are still very dense and the reverse shock generated in the outer shells of the supernova remnant (SNR) has not heated the inner ejecta yet. 
Thus, photo-electric absorption from this  metal-rich material can hide the X-ray emission of a hypothetical compact object. During the last years, many works \citep{omp15,alf18a,erl18,pbg20} investigated the upper limit on the luminosity of the putative compact leftover in various wavelengths, considering the case of a neutron star (NS) emitting thermal (black-body) radiation and obscured by the cold ejecta and/or dust. However, the lack of strong constraints on the absorption pattern in the internal area of SN87A prevented either to further constrain the luminosity of the putative NS or to make predictions about its future detectability.

On the other side, the X-ray emission from a young NS may include a significant nonthermal component: the synchrotron radiation arising from the pulsar wind nebula (PWN) associated with the rotating NS. Recently, {\it ALMA} images showed a {\it blob} structure whose emission is somehow compatible with the radio emission of a PWN  \citep{cmg19}. However, the authors themselves warned that this blob could be associated with other physical processes, e.g. heating due to $^{44}$Ti decay. It is natural, then, to look for the high-energy counterpart of the synchrotron radiation in the X-ray band. 

In this letter, we report on the analysis of observations of SN87A performed between 2012 and 2014 by {\it Chandra} and {\it NuSTAR}. We also take advantage of the state-of-the-art MHD simulation from \citet{oon20} (hereafter Or20) to reconstruct the absorption pattern within SN87A, and link it to the observed spectra. We provide a single model describing the emission in these years from 0.5 to 20 keV, isolating the synchrotron radiation arising from the remnant.
We then discuss the possible origin of such emission and show how the presence of a PWN appears to be the most likely scenario.

\section{X-ray data analysis} \label{sec:x-ray_data}

We used data collected in 2012, 2013 and 2014 with {\it Chandra} ACIS-S and {\it NuSTAR} CZT (FPMA and FPMB). We reprocessed {\it Chandra} and {\it NuSTAR} data with the standard pipelines available within CIAO v4.12.2 and NuSTARDAS, respectively. For details on the observations and the data reduction see Appendix \ref{sect:appendix:tab_obs}.

\begin{figure*}
\centering
\begin{minipage}{0.32\textwidth}
  \includegraphics[width=\textwidth]{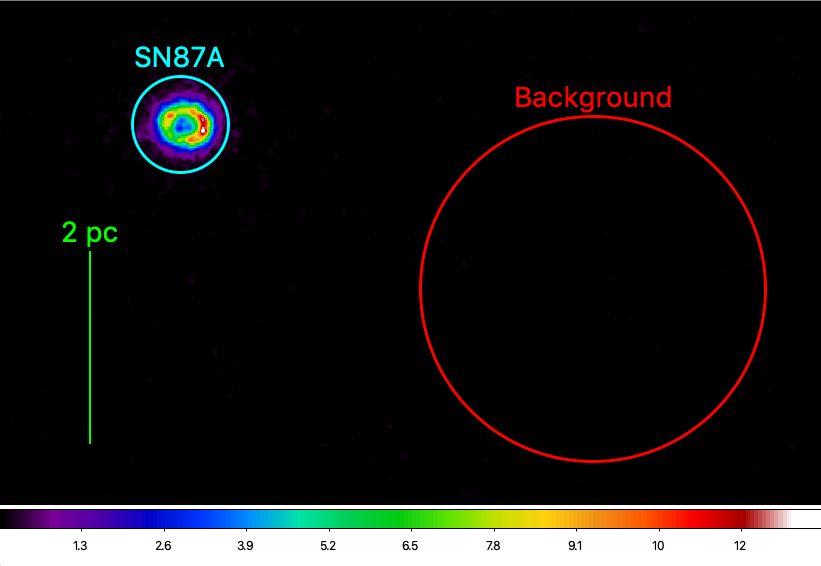}
  \end{minipage}
  \begin{minipage}{0.3\textwidth}
  \includegraphics[width=\textwidth]{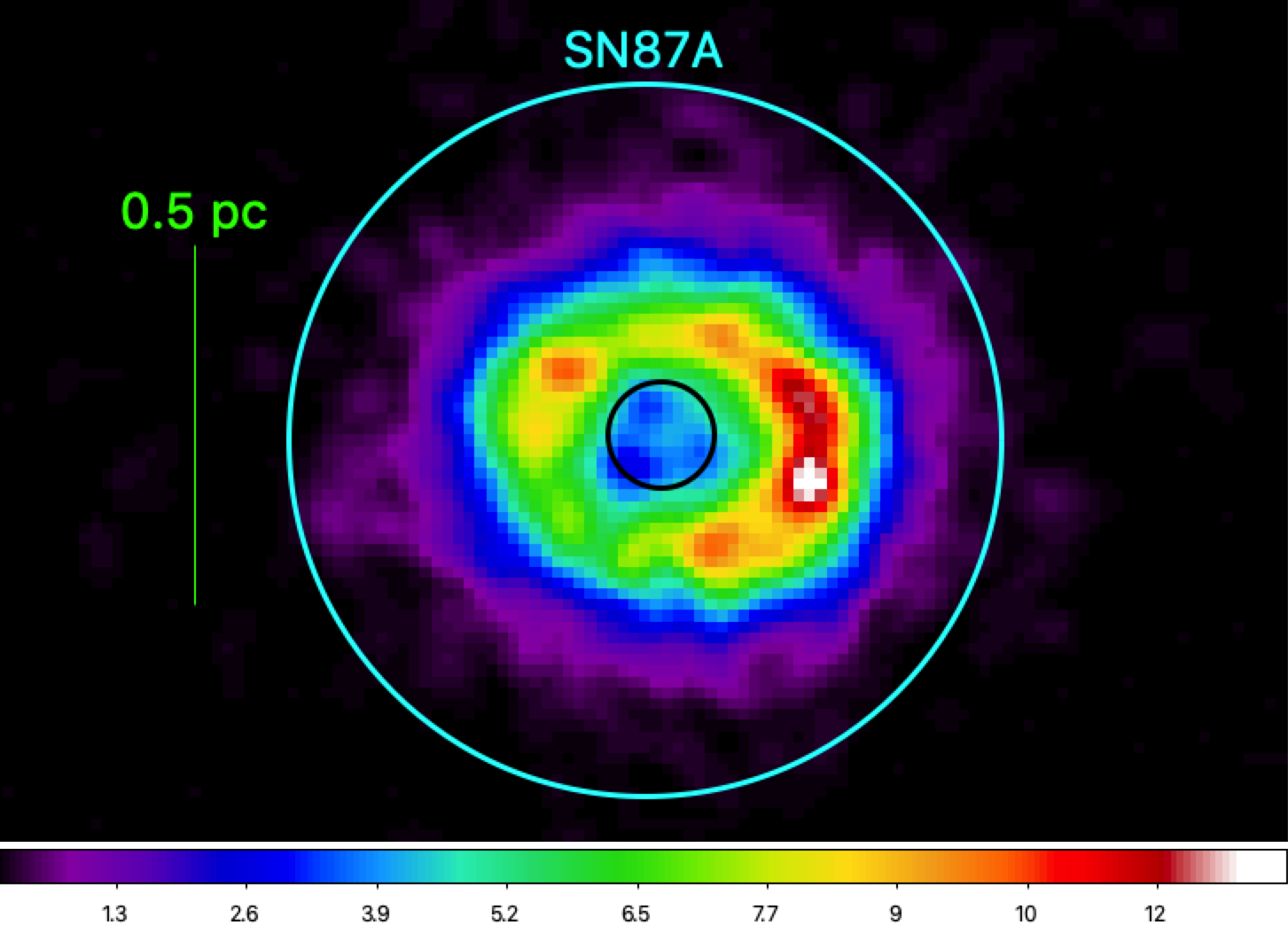}
  \end{minipage}
  \begin{minipage}{0.32\textwidth}
  \includegraphics[width=\textwidth]{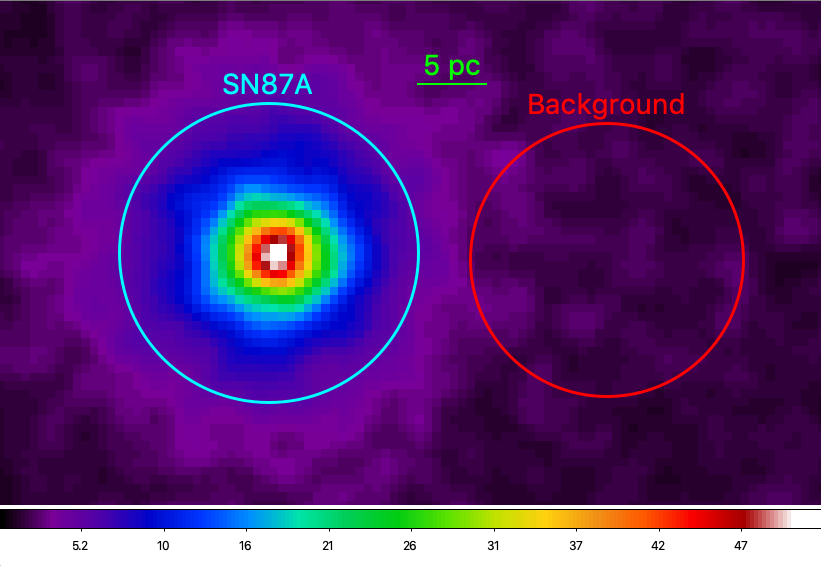}
  \end{minipage}
    \caption{ {\it Chandra} and {\it NuSTAR} count images of SN87A. All the images are smoothed with a 1.5$\sigma$-gaussian. \emph{Left panel}. {\it Chandra} image in the 0.1-8 keV. The cyan and the red circular regions mark the $2''$ region used to extract the source and the corresponding background spectra for the {\it Chandra} data, respectively. \emph{Central panel}. Closeup of the left panel. The $0.3''$ radius black ring marks the faint central region of SN87A. \emph{Right panel.} {\it NuSTAR} image in the 3-30 keV band. The cyan and red regions identify the $43''$ region used to extract the source spectra and the corresponding background for the {\it NuSTAR} data, respectively. A navigable 3D graphic showing the PWN position in the remnant interior and the comparison with the observations is available at the link \url{https://skfb.ly/6XZIU.}} 
    \label{fig:image_obs}
\end{figure*}

We extracted spectra from a circular region centered at $\alpha = 5^h 35^m 28^s$ and  $\delta = -69^{\circ}16' 10''$ with a radius of $2''$ and $43''$ for {\it Chandra} and {\it NuSTAR} data, respectively. These regions enclose all the remnant and were chosen taking into account the PSFs of the different telescopes (see Fig.~\ref{fig:image_obs}).

We simultaneously analyzed {\it Chandra} and {\it NuSTAR} spectra for each year considered by adopting a model composed by a galactic absorption component ({\it TBabs} model in XSPEC), two optically thin isothermal components in non-equilibrium of ionization ({\it vnei} model) and a constant factor which takes into account cross-calibration between different detectors. The column density n$_{\rm{H}}$ is fixed to $2.35 \times 10^{21} \, \mathrm{cm}^{-2}$ \citep{pzb06}.
We found cross-calibration factors $<2\%$ within different \emph{Chandra} data sets of the same year, and $<8\%$ between {\it Chandra} and {\it NuSTAR}, compatible with the characteristic corrections between {\it NuSTAR} detectors and other telescopes. Temperatures, emission measure, ionization age and abundances of O, Ne, Mg, Si and S were left free to vary in the fitting process. All other abundances were kept fixed to those found by \cite{zmd09}. We report no significant variations in the chemical abundances in the time range considered (2012 to 2014).

\begin{figure*}[!ht]
 \begin{minipage}{0.5\textwidth}
  \includegraphics[width=.7\textwidth,angle=270]{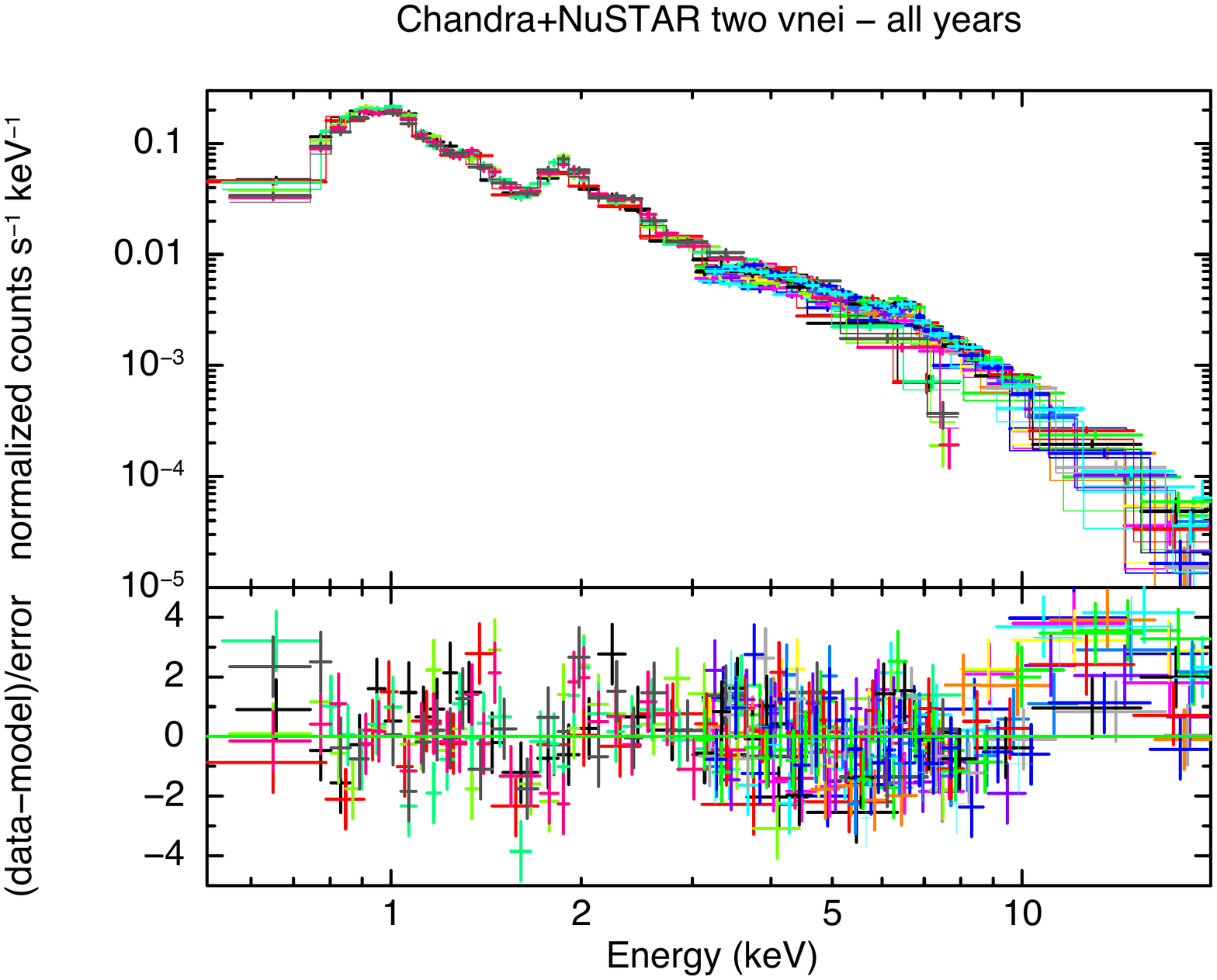}
  \end{minipage}
  \hfill
  \begin{minipage}{0.5\textwidth}  
  \includegraphics[width=.7\textwidth,angle=270]{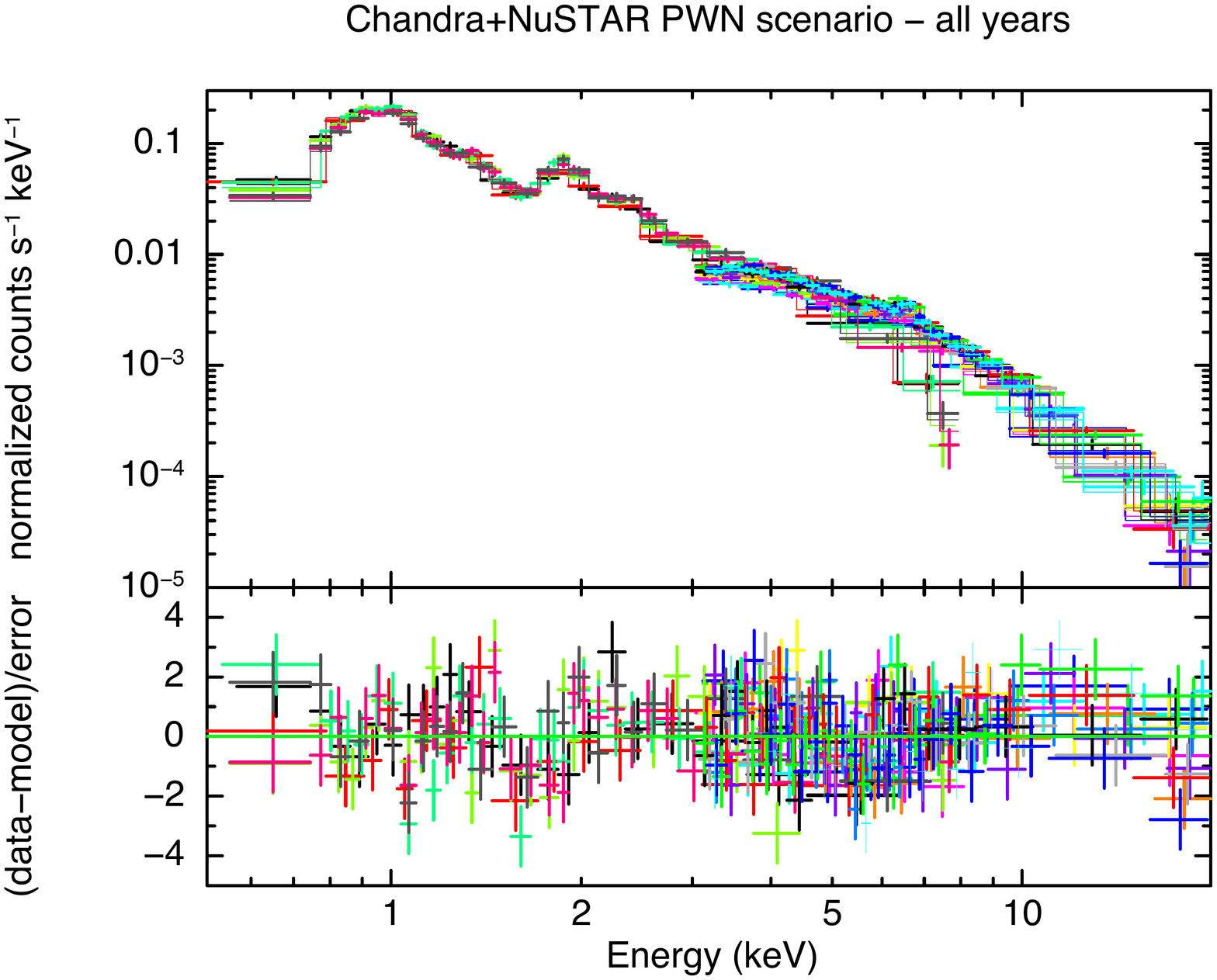}
  \end{minipage}
\caption{Spectra extracted in the $0.5-20$ keV energy band from {\it Chandra} and {\it NuSTAR} in various years with the corresponding best-fit model and residuals. A different color is associated with each of the twenty data set. On the left, the best-fit model is composed by two thermal components. On the right, the best-fit model also takes into account the emission coming from a heavily absorbed PWN. The spectra have been rebinned for presentation purposes.}
    \label{fig:chandra+nu_spectra}
\end{figure*}

In agreement with previous works \citep[Or20]{omp15, mob19}, we found that the soft ($0.5-8$ keV) X-ray emission consists of thermal X-rays originating in the shocked CSM. Our best-fit values of temperatures, $kT$ and ionization parameters, $\tau$, are compatible with previous measures \citep{zmd09}. However, we found strong residuals in all {\it NuSTAR} spectra at energies $> 10$ keV, clearly showing that an additional component must be added to the model to properly describe the hard X-ray emission (see \citealt{rzb15}). The best-fit model and the residuals are shown in the left panel of Fig.~\ref{fig:chandra+nu_spectra}. This additional component cannot be associated with thermal emission from the shocked plasma, since otherwise unrealistically high temperatures would be necessary ($kT\sim20$ keV), one order of magnitude higher than the maximum electron temperature predicted for SN 1987A (\citealt{omp15}) and never observed in SNRs.

If the emission is unlikely to be thermal, it is plausible to suppose that it is nonthermal. This nonthermal emission may arise from a possible compact object, most likely a NS embedded in its PWN (hereafter PWN87A), which emits synchrotron radiation. The main issue in tackling this scenario is to isolate the radiation coming from this object.

We can estimate the absorbing power of cold ejecta surrounding the PWN by taking into account the abundance and density pattern along the line of sight provided by the MHD model by Or20 (see Appendix \ref{sec:ray_tracing}) in the considered years. To include this information in the spectral analysis, we added an absorbed power-law modeled with {\it vphabs} within XSPEC. Since its parameters were derived from the model they are not free to vary in the fit.
With this additional component, we obtained a very good description of the observed spectra over the whole $0.5-20$ keV energy band (right panel of Fig.~\ref{fig:chandra+nu_spectra}). The best-fit value of photon index and normalization of the absorbed power-law do not change significantly between the three years considered. We then fitted simultaneously the 2012, 2013 and 2014 spectra in order to decrease the uncertainties in the best-fit parameters of the power-law component. Temperatures, ionization parameters and normalization of thermal emission  were free to change in time, while normalization and photon index of nonthermal emission (and chemical abundances) were left free to vary in the fitting procedure, but forced to be the same over the three years. 

The resulting PWN best-fit photon index is $\Gamma = 2.5^{+0.3}_{-0.4} $ and the X-ray luminosity in the $1-10$ keV band is $L_{1-10} = (2.6 \pm 1.4) \times 10^{35}$ erg/s (Table \ref{tab:fit_whole}). Our best fit values are compatible with typical values found for PWNe (Sect. \ref{sect:pwnvsdsa}).

We found an increase in flux in the $8-20$ keV band between 2012 and 2014 (as reported by \citealt{rzb15}). We point out that the emission from the PWN is consistent with being constant, and that the increase in the hard X-ray flux is related to the steadily growing thermal emission arising from the  interaction of the remnant with the HII region.

Because of the high ejecta absorption, the PWN flux is strongly suppressed below 8 keV. In particular, we found that the PWN emission is much dimmer ($\sim 4\%$) than the  thermal emission observed by \emph{Chandra} in the faint central region of SN87A (identified by a circular region with radius $R=0.3''$, well inside the bright X-ray emitting ring, see central panel in Fig.  \ref{fig:image_obs}).
 
As an alternative scenario, we also considered the case in which the hard X-ray emission is associated with synchrotron radiation due to DSA occurring in the outer layers of SN87A. We then removed the cold ejecta absorption. The best-fit photon index is $\Gamma_{\rm{DSA}} = 2.0 \pm 0.4$, indicating lower steepness than the PWN one but still consistent with it. Analogously to what we obtained for the PWN scenario, we found that photon index and normalization of the nonthermal emission do not change significantly from 2012 to 2014. We then repeated the simultaneous fit of the data from all years obtaining a good description of the observed spectra ($\chi_{\rm{DSA}}^2=1457$ with 1221 d.o.f. to be compared with $\chi^2_{\rm{PWN}}=1442$ with 1221 d.o.f. of the PWN scenario).

\begin{table*}
    \caption{Best-fit parameters of the model adopted to describe {\it Chandra} and {\it NuSTAR} observations performed in 2012, 2013 and 2014. Chemical abundances and power law parameters are kept constant along the various years. Uncertainties are at 90\% confidence level.}
    \centering
    \begin{tabular}{c|c|c|c|c}
    \hline\hline
    Component& Parameter & 2012 & 2013 & 2014 \\
    \hline
    TBabs & n$_{\rm{H}}$ (10$^{22}$ cm$^{-2}$) & \multicolumn{3}{c}{0.235 (fixed)} \\
    \hline
    & kT (keV) & 2.85$_{-0.07}^{+0.08}$ & 2.87$\pm 0.07$ & 2.85$\pm 0.05$\\
    & O  & \multicolumn{3}{c}{0.32$_{-0.03}^{+0.04}$} \\
    & Ne & \multicolumn{3}{c}{0.57$ \pm 0.02$}\\
    vnei & Mg & \multicolumn{3}{c}{0.36 $\pm 0.03$}\\
    & Si & \multicolumn{3}{c}{0.39$_{-0.02}^{+0.03}$}\\
    & S  & \multicolumn{3}{c}{0.65$ \pm 0.06$}\\
    & $\tau$ (10$^{11}$ s/cm$^3$)& 2.1$_{-0.5}^{+0.9}$& 1.7$\pm 0.3$& 1.8$_{-0.2}^{+0.4}$\\
    & EM ($10^{58} \mathrm{cm}^{-3}$) & 7.0$ \pm 0.3$& 7.3$\pm 0.3$ & 8.2$_{-0.2}^{+0.3}$ \\
    \hline
    & kT (keV)& 0.60$\pm 0.04$& 0.65$_{-0.03}^{+0.04}$& 0.63$_{-0.3}^{+0.4}$\\
    vnei & $\tau$ (10$^{11}$ s/cm$^3$) & 1.8$_{-0.3}^{+0.5}$& 1.5$\pm{0.3}$& 1.7$\pm 0.3$ \\
    & EM ($10^{58} \mathrm{cm}^{-3}$) & 31$_{-3}^{+2}$ & 26$\pm 2$& 28$\pm 0.2$ \\
    \hline
    pow& Photon index $\Gamma$ & \multicolumn{3}{c}{2.5$_{-0.4}^{+0.3}$} \\
    \hline
    \multicolumn{2}{c|}{L$^{\rm{pwn}}_{0.5-8}$ (10$^{35}$ erg/s)} & \multicolumn{3}{c}{$4.1_{-2.8}^{+4}$} \\
    \multicolumn{2}{c|}{L$^{\rm{pwn}}_{1-10}$ (10$^{35}$ erg/s)} & \multicolumn{3}{c}{ $2.6 \pm 1.4 $} \\
    \multicolumn{2}{c|}{L$^{\rm{pwn}}_{10-20}$ (10$^{35}$ erg/s)}& \multicolumn{3}{c}{0.32$_{-0.01}^{+0.02}$}\\
    \hline
    \multicolumn{2}{c|}{Flux$_{0.5-8}$ (10$^{13}$ erg/s/cm$^2$)}& 93$\pm 2$ & 92$_{-2}^{+4}$& 95$_{-1}^{+2}$\\
    \multicolumn{2}{c|}{Flux$_{10-20}$ (10$^{13}$ erg/s/cm$^2$)}&1.5$\pm 0.2$ &1.6$_{-0.1}^{+0.6}$ &1.6$_{-0.1}^{+0.3}$\\
    \hline
    \multicolumn{2}{c|}{$\chi^2$ (d.o.f.)} & \multicolumn{3}{c}{1442 (1221)} \\
    \end{tabular}
    \label{tab:fit_whole}
\end{table*}

Since from a statistical point of view an improvement of only 1\% is not enough to favour the PWN case over the DSA one, in Sect. \ref{sect:pwnvsdsa} we present a comparison of the physical implications in both cases.

\section{Discussion} \label{sect:pwnvsdsa}
In Sect. \ref{sec:x-ray_data} we showed that a nonthermal component is needed to properly fit the {\it NuSTAR} data in the $10-20$ keV band. As already mentioned, the possible origin of the physical mechanism responsible for such emission is twofold: either DSA or emission from an heavily absorbed PWN. Under a spectroscopic perspective, the only difference between the two scenarios is the presence of an additional absorption component in the PWN case, the \emph{vphabs} model, because of the presence of cold ejecta surrounding the putative compact object. The heavy absorption leads to a negligible contribution of the power-law at energies below 6 keV and thus to a steeper $\Gamma$. However, the photon index values in the two scenarios are compatible with each other taking into account the 90\% confidence error bars. Moreover, the very similar values of $\chi^2$ does not allow us to exclude one of the two possible emission mechanisms under a merely statistical point of view.

In the DSA scenario, the flux is expected to vary with time in the same way both in the X-ray and radio bands. The synchrotron radio flux of SN87A increased by $\sim15\%$ between 2012 and 2014 (\citealt{cgn18}). Under the DSA hypothesis, this woud be at odd with our findings, since we observe a steady X-ray synchrotron emission, and a $15\%$ increase of the nonthermal X-ray flux is discouraged at the 90\% confidence level. However, we note that the DSA scenario could allow for spectral variations in time that would make X-ray variations less predictable from radio variations.

To further investigate the emission nature in the DSA scenario, we replaced the power-law component with the XSPEC {\it srcut} model, which describes synchrotron emission from an exponentially cut off power-law distribution of electrons in the assumption of homogeneous magnetic field \citep{rey98}. For each of the three years considered, we constrained the normalization of the \emph{srcut} component (i.e., its flux at 1 GHz, $S_{1GHz}$), from the corresponding values observed at 9 GHz by \citet{cgn18}, taking into account the radio spectral index of SN87A ($\alpha=0.74$,  \citealt{zsn13}). We fixed $S_{1GHz}$ and $\alpha$ in the {\it srcut} model and left the break frequency, $\nu_b$, free to vary. The radio-to-X-rays spectral index $\alpha$ can be assumed to be constant since the synchrotron cooling time is much longer than the age of the system, thus no significant cooling is expected (i.e. $\tau_{\rm{sync}}\sim125$ yr $(E/10~{\rm TeV})^{-1}~(B/100~\mu{\rm G})^{-2}$).
We obtained $h\nu_b=2.4_{-0.4}^{+0.3}$ keV, $h\nu_b=2.3_{-0.4}^{+0.3}$ keV and $h\nu_b= 1.7_{-0.2}^{+0.4}$ keV, in 2012, 2013 and 2014 respectively. 
The break energy is compatible with 2 keV in the three years considered, though the general trend seems to point towards a decrease with time.

The synchrotron emission of electrons peaks at energy $h\nu = 1.8 \times 10^4 E^2_{100} B_{100}$, where $B_{100}$ is the magnetic field in units of 100 $\mu G$ and $E_{100}$ is the electron energy in units of 100 TeV. Considering our break energy $h\nu_b\sim2$ keV, we obtain values of $E_{100}$ spanning from $\sim 0.14$ to $\sim 0.33$, for $B_{100}$ ranging from 6 to 1. This maximum electron energy seems to be quite high, especially considering the relatively low shock speed. In fact, the synchrotron radio emission originates in the HII region \citep{zsn13,cgn18,omp19}, where the shock velocity is of only 2000 km/s \citep[Or20]{cgn18}.

In the DSA scenario, we can estimate the acceleration time scale as \citep{pmb06} $\tau_{\rm{acc}}=124 \eta B^{-1}_{100} V_s^{-2} E_{100} \frac{4}{3}$ yr, where $\eta$ is the Bohm factor and $V_s$ is the shock velocity. In the hypothesis of maximum efficiency ($\eta = 1$) and a standard magnetic field $B_{100} =1$, with $V_s$ = 2000 km/s, we would need $\tau_{acc}\sim390$ yr to accelerate the X-ray emitting electrons up to the observed maximum energy, i.e., much more than the age of SN87A. The observed maximum energy can be obtained in 25 yr only by assuming that the downstream magnetic field is amplified by the SN87A slow shock up to $\sim600$ $\mu$G (in this case the maximum electron energy would be of $\sim14$ TeV), and only assuming that the acceleration proceeds at the Bohm limit. The aforementioned issues (steady synchrotron flux and extremely large electron energy in a relatively slow shock) concur in making the DSA scenario unconvincing.

On the other hand, the PWN scenario has a strong physical motivation, as we show below. In the absence of a direct identification of the compact object eventually powering PWN87A, the only possible way to constrain its properties is to use the X-ray luminosity obtained in our analysis to locate the putative PWN within the PWNe population.

The properties of a generic PWN can be associated to those of its progenitor SNR and surrounding ISM introducing the characteristic time and luminosity scales \citep{tm99}:

\begin{eqnarray}
 t_{\rm{ch}} &=& E_{\rm{sn}}^{-1/2} M_{\rm{ej}}^{5/6} \rho_{\rm{ism}}^{-1/3}\,, \label{eq:chscales1} \\ 
 L_{\rm{ch}} &=& E_{\rm{sn}}/ t_{\rm{ch}}\,, \label{eq:chscales2}\
\end{eqnarray}
where $E_{\rm{sn}}$ is the supernova explosion energy, usually assumed to be $10^{51}$ erg, $M_{\rm{ej}}$ is the mass in the SNR ejecta and $\rho_{\rm{ism}}$ the mass density of the ISM.
These last parameters have been considered to vary uniformly in: $M_{\rm{ej}} \in [5-20] \, \rm{M_\odot}$ \citep{sec09} and $n_{\rm{ism}}\in[0.01-10]$ cm$^{-3}$ \citep{ber87,mpp97,lbw10,bp10,asv14}, where $\rho_{\rm{ism}}=m_p n_{\rm{ism}}$  and $m_p$ is the proton mass. For the pulsar population the best choice is to consider young $\gamma-$ray emitting pulsars \citep{wr11,jsk20}, better suited for describing pulsars powering PWNe than the old radio-emitting ones \citep{kr06}. 

Considering the pulsar parameters from that  population (namely the initial spin-down time $\tau_0$ and luminosity $L_0$), with the choice of the canonical dipole braking index $n=3$, the PWNe population can be then constructed\footnote{The PWNe population is liable of changes in the parameters plane, especially if considering a different braking index value than the  standard  dipole  one  (\citealt{Parthasarathy:2020}.)} scaling time and luminosity with the characteristic ones defined in Eq.~\ref{eq:chscales1}-\ref{eq:chscales2}. In the $(\tau_0/t_{\rm{ch}},\,L_0/L_{\rm{ch}})$ plane the PWNe population appears as an ellipsoidal surface, where each point corresponds to various physical sources with different combinations of $\tau_0, L_0, M_{\rm{ej}}$ and $\rho_{\rm{ism}}$. 
We use this PWNe population to discuss the possible location of PWN87A, by scaling the observed X-ray luminosity for the corresponding characteristic quantities.
In particular, the ejecta mass and kinetic energy are $18\rm{M_\odot}$ and $2\times10^{51}$ erg, respectively (Or20, and references therein). Given the complex structure of the remnant, the density of the material in which the ejecta expand shows large inhomogeneities, varying from $\sim 0.1$ cm$^{-3}$ in the pre-shock blue supergiant wind,  $\sim 100$ cm$^{-3}$ in the HII region, up to $10^3-10^4$ cm$^{-3}$ in the dense ring \citep[Or20]{sck05}. 
We consider a value of $\sim 100$ cm$^{-3}$, representative of the equatorial zone  of the HII region. 
Thus the two scaling are:
\begin{eqnarray}
t^{87\rm{A}}_{\rm{ch}}  &=&  
807.6\,\rm{yr} \,\left( \frac{E_{\rm{sn}}}{2\times10^{51} \,\rm{erg}} \right)^{-1/2}
 \left( \frac{M_{\rm{ej}}}{ 18 \rm{M_\odot}} \right)^{5/6}		\nonumber\\
& \quad & \left( \frac{\rho_{\rm{ism}}}{ 100 m_p \rm{cm}^{-3}} \right)^{-1/3}\,,\\
 L^{87\rm{A}}_{\rm{ch}} &=& 7.86\times 10^{40}\,\rm{erg/s} \left( \frac{E_{\rm{sn}}}{2\times10^{51} \,\rm{erg}} \right)
 \,.
\end{eqnarray}\label{eq:chscales_1987A}

Taking into account the ejecta absorption, as calculated from our MHD simulation, we derive from the spectral analysis the unabsorbed X-ray luminosity of the central source, finding L$_{X}^{\rm{pwn}} = 4.1^{+4}_{-2.8} \times 10^{35}$ erg/s in the $0.5-8$ keV band, considering a distance to the source of $51.4$ kpc.
The X-ray luminosity is related to the spin-down power $\dot{E}$ through the conversion efficiency parameter $\eta_{X}=L_{X}/\dot{E}$, that shows very large variations within the population, from $\sim10^{-5}$ to $\sim10^{-1}$ \citep{kp08}.

\begin{figure*}[!ht]
\begin{minipage}{0.5\textwidth}
  \includegraphics[width=.9\textwidth]{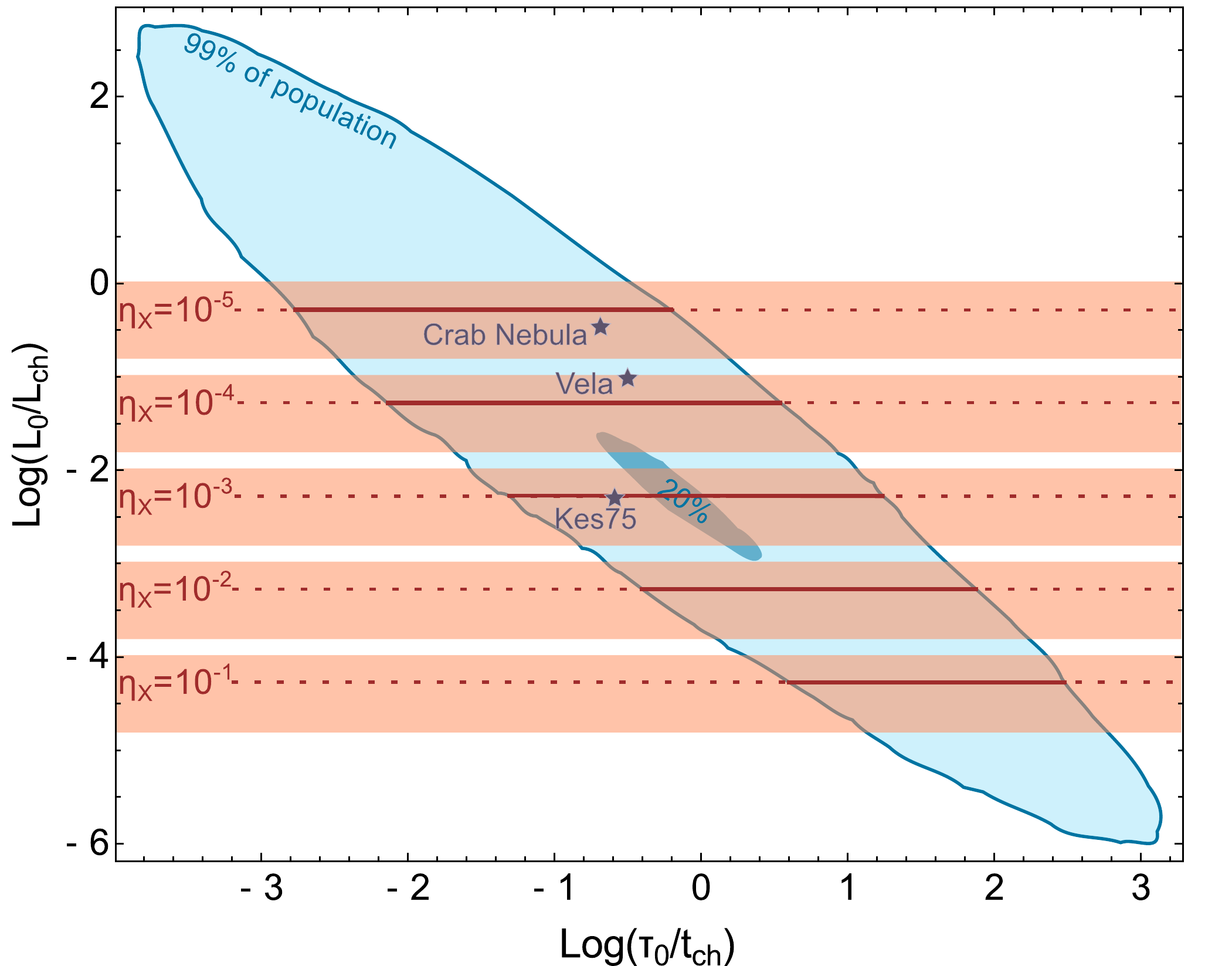}
  \end{minipage}
  \hfill
  \begin{minipage}{0.5\textwidth}
  \includegraphics[width=.9\textwidth]{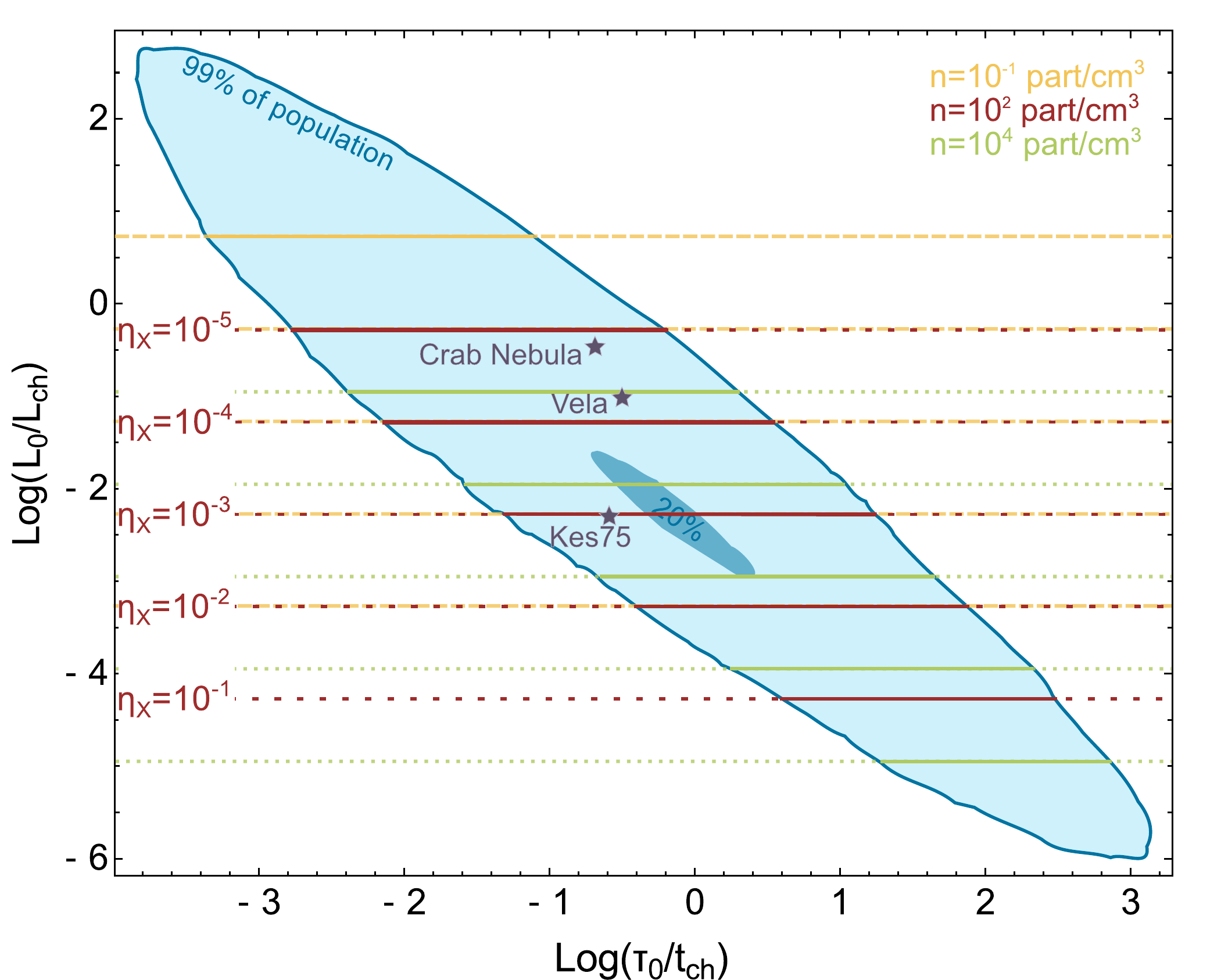}
  \end{minipage}
  \caption{\emph{Left panel}. Positioning of the putative PWN87A (brown lines) within the PWNe population (in light-blue color) in the case of the representative density 100 cm$^{-3}$ of the HII region. The PWNe population is represented with contours enclosing the 99\% of the entire population and the 20\% (to better mark the center). 
    Different possible positions of the PWN have been obtained using the detected X-ray flux in the 0.5-8 keV band (with the best fit value shown as a solid-dashed line and the entire range of variation as a shaded area) and considering the widest possible variation for the X-ray efficiency $\eta_X$ \citep{kp08}. \emph{Right panel}. Variation of the predicted location of the PWN87A within the PWNe population, for different values of $\eta_X$, considering an extreme variation of the ambient density, from $0.1$ cm$^{-3}$ (representative of the blue giant wind, in yellow) to $10^4$ cm$^{-3}$ (the maximum density in the dense ring in the HII region, in green).}
\label{fig:pwnpop}
\end{figure*}
To maintain our analysis as general as possible, here we have considered the entire range of measured $\eta_X$. In Fig.~\ref{fig:pwnpop} we show the population of PWNe as determined associating the $\gamma-$ray emitting pulsars with the discussed ranges of parameters for SNRs and ISM (in light blue). The positioning of the putative PWN in SN87A for varius $\eta_X$ is shown as brown lines that intersect horizontally the distribution.
The Crab nebula, the Vela nebula and the PWN in Kes 75 are shown for comparison.
As it can be easily seen, all the different possibilities lead to a location of PWN87A fully compatible with the population.
Assuming an ambient density of 100 cm$^{-3}$ might not be correct for SN 1987A, because of the complex density distribution. Therefore, we relaxed this assumption and, in the right panel of Fig.~\ref{fig:pwnpop}, we show that the putative positions for varying $\eta_X$ are perfectly consistent with the PWN population, even considering the two extreme density values for the surrounding medium ($0.1-10^4$ cm$^{-3}$, \citealt[Or20]{sck05}).

We verified that our estimate of the PWN luminosity in the X-rays is consistent with the radio luminosity derived by \citet{cmg19} under the (very reasonable) assumption that the low energy break frequency of the synchrotron radiation is\footnote{By assuming a radio photon index $\Gamma_R=1.5$ and an X-ray photon index $\Gamma_X=2.5$.} $\nu_b\gtrsim 10^{12}$ Hz, while we expect that the high absorption of the cold ejecta prevents the detectability of the PWN emission in the optical band.

An estimate (purely based on a statistical argument) of the most probable spin-down time for each $\eta_X$ can be determined by combining the PWNe probability distribution with the estimates of the pulsar spin-down luminosity (Table~\ref{tab:pwnVal}, with $n_{\rm{ism}}=100$ cm$^{-3}$). 
\begin{table}
    \caption{Summary of the most probable values for the pulsar spin-down time for different values of the X-ray efficiency $\eta_X$, and the associated spin-down luminosity (from the best-fit value). All values are given considering $n_{\rm{ism}}=100$ cm$^{-3}$.}
    \label{tab:pwnVal}
\vspace{0.4cm}
    \centering
    \begin{tabular}{c|c|c}
    \hline\hline
    $\eta_X$& L$_{\rm{0}}$ (erg/s)& $\tau_0$ (yr) \\
    \hline
    $10^{-5}$&   $4.1 \times 10^{40}$  &  18 \\
    $10^{-4}$&   $4.1 \times 10^{39}$  &  110\\
    $10^{-3}$&   $4.1 \times 10^{38}$  &  480 \\
    $10^{-2}$&   $4.1 \times 10^{37}$  &  3160\\
    $10^{-1}$&   $4.1 \times 10^{36}$  &  25000\\
    \hline
    \end{tabular}
\end{table}

In conclusion, a PWN seems to be the most likely source of the synchrotron radiation in hard X-rays of SN87A, even though the DSA scenario cannot be firmly excluded. A more conclusive way to discern between the two scenarios will be provided by future observations. An increase in the hard ($>10$ keV) X-ray flux, similar to that observed in radio would be easily detectable with {\it NuSTAR}, thus supporting the DSA scenario. A decrease would support the PWN scenario. On the other hand, the rapid ejecta expansion and rarefaction will reduce the soft X-ray absorption of an inner source. In particular, from our MHD simulation, we estimate that the PWN will become detectable by \emph{Chandra} and/or Lynx\footnote{https://www.lynxobservatory.com} (Appendix \ref{app:lynx}) in the 2030s, definitely confirming the PWN scenario. 

With the present data and based on our findings, the PWN scenario seems the most likely (and appealing) to account for the nonthermal X-ray emission that we have detected.

\software{CIAO \citep{ciao}, HEASOFT\footnote{https://heasarc.gsfc.nasa.gov/docs/software/heasoft/}, XSPEC \citep{arn96}, NuSTARDAS\footnote{https://heasarc.gsfc.nasa.gov/docs/nustar/analysis/nustar\_swguide.pdf}}

\facilities{{\it Chandra}\footnote{https://cxc.harvard.edu/index.html}, {\it NuSTAR} \citep{NuSTAR}}

\section*{Acknowledgements}
We thank the anonymous referee for the helpful comments. EG, MM, SO, and FB acknowledge financial contribution from the INAF mainstream program. MM, EG and SO acknowledges contribution from {\it NuSTAR} (NARO18) in the framework of the ASI - INAF agreeement I/037/12/0. This work is supported by JSPS Grants-in-Aid for Scientific Research “KAKENHI A” Grant Numbers JP19H00693. SN, MO, AD wish to acknowledge the support from the Program of Interdisciplinary Theoretical \& Mathematical Sciences (iTHEMS) at RIKEN, and the support from Pioneering Program of RIKEN for Evolution of Matter in the Universe (r-EMU).

\bibliography{references}
\bibliographystyle{aasjournal}

\newpage
\clearpage
\appendix
\section{Data reduction}
\label{sect:appendix:tab_obs}

{\it Chandra} data were reprocessed with the CIAO v4.12.2 software, using CALDB 4.9.2. We reduced the data through the task {\it chandra\_repro} and we extracted the ACIS-S spectra by using the tool {\it specextract} which also provided the corresponding arf and rmf files. 

{\it NuSTAR} data were reprocessed with the standard pipelines provided by the {\it NuSTAR} data analysis software NuSTARDAS\footnote{https://heasarc.gsfc.nasa.gov/docs/nustar/analysis/nustar\_swguide.pdf} by using {\it nupipeline} and {\it nuproducts}. Details of the observations are reported in Table \ref{tab:obs}.

\begin{table*}[!ht]
    \centering
     \caption{Summary of the main characteristics of the analyzed observations.}
    \begin{tabular}{c|c|c|c|c}
     Telescope &OBS ID& PI& Date (yr/month/day) & Exposure time (ks)  \\
     \hline\hline
     & 13735& Burrows& 2012/03/28 &48\\
     & 14417& Burrows& 2012/04/01&27\\
     {\it Chandra}&14697& Burrows& 2013/03/21 &68\\
     &14698& Burrows& 2013/09/28 &68 \\
     & 15809& Burrows & 2014/03/19 &70\\ 
     & 15810& Burrows& 2014/09/20 &48\\
     \hline
     & 40001014003& Harrison& 2012/09/08&136\\
     & 40001014004& Harrison& 2012/09/11&200\\
     & 40001014007& Harrison& 2012/10/21&200\\
     {\it NuSTAR}& 40001014013& Harrison& 2013/06/29 &473\\
     & 40001014018& Harrison& 2014/06/15&200\\
     & 40001014020& Harrison& 2014/06/19&275\\
     & 40001014023& Harrison& 2014/08/01&428\\
    \end{tabular}
    \label{tab:obs}
\end{table*}

Spectral analysis has been performed with XSPEC (v12.11.1, \citep{arn96}) in the $0.5-8$ keV and $3-20$ keV bands for the \emph{Chandra} and \emph{NuSTAR} data, respectively. All spectra were rebinned adopting the optimal binning procedure described in \citet{kb16}, and the background spectrum to be subtracted was extracted from a nearby region immediately outside of the source. We verified that our results are not affected by the choice of the background regions.

\section{X-ray absorption from cold ejecta} \label{sec:ray_tracing}

We used the 3D MHD simulation of SN87A by Or20 to estimate the absorption pattern of the cold ejecta. The simulation reproduces most of the features observed in the remnant of SN87A in various spectral bands, and links the SNR with the properties of the asymmetric parent SN explosion \citep{onf20} and with the nature of its progenitor star \citep{uru18}. The model provides all relevant physical quantities in each cell of the 3D spatial domain (MHD variables, plasma composition of the CSM and of the ejecta vs time). The most notable quantities for our purposes are: the electron temperature, the ion density for many chemical species ($^{1}$H, $^{3}$He, $^{4}$He, $^{12}$C, $^{14}$N, $^{16}$O, $^{20}$Ne, $^{24}$Mg, $^{28}$Si, $^{32}$S, $^{36}$Ar, $^{40}$Ca, $^{44}$Ti, $^{48}$Cr, $^{52}$Fe, $^{54}$Fe, $^{56}$Ni, and the decay products, including $^{56}$Fe), and the ionization age. The model also predicts that the putative NS relic of the supernova explosion has received a kick towards the observer in the north with a lower limit to the kick velocity of $\approx 300$~km/s, resulting from a highly asymmetric explosion \citep[Or20]{onf20}. 

For the present study, we assumed a slightly higher kick velocity of 500 km/s. We checked that the results do not change significantly for values ranging between 300 and 700 km/s. Given the kick velocity and the direction of motion of the NS as predicted by the model \citep[Or20]{onf20}, and orienting the modeled remnant as it is observed in the plane of the sky (Or20), we established the position of the NS in the 3D spatial domain of the simulation for each year analyzed in this work. 
Considering that the extension of the putative radio PWN is of the order of $<1000$ AU \citep{cmg19}, that the X-ray PWN is expected to be smaller than its radio counterpart, and that the spatial resolution of our MHD simulations is of $\approx 180$ AU, we can consider the central source as point-like in our procedure.

We reconstructed the absorption pattern encountered by the synchrotron X-ray emission of the putative PWN through the subsequent absorbing layers along the line of sight. The physical effect responsible for the absorption is the photo-electric effect, since the material surrounding the considered source is cold ($T < 100$ K). We extracted values of temperature, column density and abundances associated to each absorbing layer of the 3D domain of the model and we included these parameters in the spectral analysis through the XSPEC photo-electric absorption model {\it vphabs}. 

For the years considered in this work, the absorption due to cold ejecta is comparable with an equivalent H column density higher than $10^{23}$ cm$^{-2}$. This indicates that potential signature of a PWN emission must be searched in the high energy part ($\ga$ 10 keV) of the X-ray spectra, less affected by absorption.

\section{Synthetic LYNX spectrum} \label{app:lynx}

\begin{figure}[!ht]
    \centering
    \includegraphics[width=0.4\textwidth,angle=270]{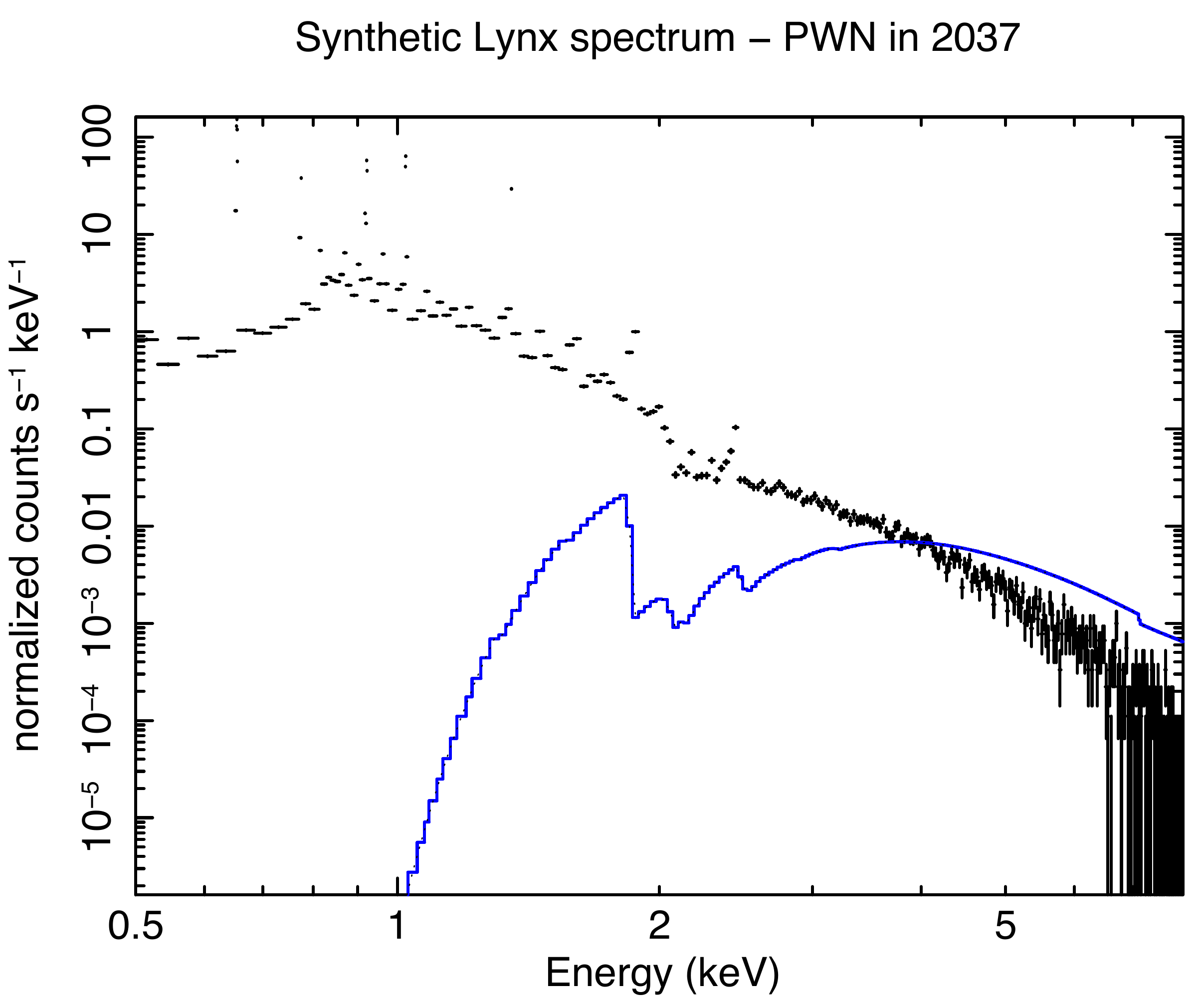}
    \caption{Black points: synthetic LYNX spectrum of the emission observed in 2018. Blue line: power-law component absorbed by the ejecta as predicted by the MHD model in 2037.}
    \label{fig:lynx_synth}
\end{figure}

 We produced a synthetic LYNX observation of SN87A, as predicted by our MHD model for year 2037. We expect the thermal emission to stay almost constant in the next 15-20 years, though ejecta contribution and/or interaction of the remnant with other inhomogeneities beyond the ring may affect our predictions. Fig. \ref{fig:lynx_synth} shows the synthetic spectrum extracted from a circular region with radius $R=0.3''$, well within the bright ring of SN87A (central panel of Fig. \ref{fig:image_obs}), assuming an exposure time of 300 ks. If we consider the emission of the PWN absorbed by the ejecta pattern that the model predicts for 2037, we notice that the resulting component would have a flux higher than that associated with thermal emission above 4 keV, thus becoming detectable in the soft x-ray band (see Fig. \ref{fig:lynx_synth}.). Further details about the future detectability of the PWN and on the thermal emission of the putative NS will be described in a forthcoming paper (Greco et al., in prep.).

\end{document}